\documentclass[12pt]{article}
\usepackage{standalone, multirow, amsmath}
\usepackage{scalerel}

\usepackage{algorithm}
\usepackage[noend]{algpseudocode}

\makeatletter
\def\BState{\State\hskip-\ALG@thistlm}
\makeatother

\usepackage{forloop}
\newcounter{ct}

\newlength\myindent
\usepackage{geometry}
\usepackage{authblk}

\begin{document}
	\title{Comprehending Real Numbers: Development of Bengali Real Number Speech Corpus}
    \date{}
    \author[*,1]{\rm Md Mahadi Hasan Nahid}
    \author[*,2]{\rm Md. Ashraful Islam }
    \author[*,2]{\rm Bishwajit Purkaystha}
    \author[*,1]{\rm Md Saiful Islam}
    \affil[*]{Department of Computer Science and Engineering}
    \affil[ ]{Shahjalal University of Science and Technology, Sylhet-3114.}
    \affil[*,1]{\textit {\{nahid-cse,saiful-cse\}@sust.edu}}
    \affil[*,2]{\textit{\{ashrafulcse.sust, bishwa420\}@gmail.com}}
    \maketitle
    \begin{abstract}
    	
Speech recognition has received a less attention in Bengali literature due to the lack of a comprehensive dataset. In this paper, we describe the development process of the first comprehensive Bengali speech dataset on real numbers. It comprehends all the possible words that may arise in uttering any Bengali real number. The corpus has ten speakers from the different regions of Bengali native people. It comprises of more than two thousands of speech samples in a total duration of closed to four hours. We also provide a deep analysis of our corpus, highlight some of the notable features of it, and finally evaluate the performances of two of the notable Bengali speech recognizers on it.

    \end{abstract}
    \section{Introductions}
    
Automatic Speech Recognition (ASR) capability of machines is very important as it reduces the cost of communication between humans and machines. The recent resurgence in ASR \cite{chan2016listen, graves2014towards} is significant and due to the large recurrent neural networks \cite{zaremba2014recurrent}. The neural networks are most useful when they have large datasets. The improvement in automatic speech recognition is not reflected in Bengali literature. This is due to a lack of comprehensive corpus in the literature. There are few corpora in the literature \cite{alam2010development}, but they are not comprehensive, meaning that they do not encompass all the possible words in a specific portion of the literature. In this paper, we address this issue by describing the development of a Bengali speech corpus which encompasses all possible words that may appear in a Bengali real number. We call our corpus ``Bengali real number speech corpus''.
\paragraph{} This corpus was developed to cover all the words that could possibly involve in any Bengali real number. The corpus comprises the speeches of several Bengali native speakers from different regions. The summary of our corpus is given in Table \ref{table:summary-of-corpus}. One of the limitations of our corpus is that its speakers are all male speakers and the standard deviation in the ages of the speakers is relatively small. The vocabulary of the corpus, mentioned in Table \ref{table:summary-of-corpus}, contains all the Bengali numbers from zero to hundred. Each number from zero to hundred in Bengali is uttered in a singular way, except for the number `45'. It is uttered as both `/p/o/y/t/a/l/l/i/sh' and `/p/o/y/ch/o/l/l/i/sh'. Therefore, 102 words in our vocabulary stand for number 0 to number 100. The details about the rest of twelve words are given in Table \ref{table:details-of-twelve-words}. All the words involved in uttering any Bengali real number fall to either the category of first 102 words or to the category of the words mentioned in Table \ref{table:details-of-twelve-words}.
\begin{table}
\centering
\caption{Bengali real number audio corpus summary}
\label{table:summary-of-corpus}
\begin{tabular}{l|l}
\multicolumn{2}{c}{} \\
\hline
Number of speakers & 10 \\
Age range & 20-23 years\\
Number of speech samples & 2302 samples \\
Duration & 3.79 hours \\
Vocabulary size & 115 words \\
Number of distinct phonemes & 30 \\
Average number of words per sample & $\approx 8$ words \\ \hline
\end{tabular}
\end{table}

\begin{table}
\centering
\caption{Special twelve words}
\label{table:details-of-twelve-words}
\begin{tabular}{c|c|l|l}
\multicolumn{4}{c}{} \\
\hline \hline
\# & Bengali word & English correspondence & Bengali utterance \\ \hline
1 & \includegraphics[width=0.05\textwidth]{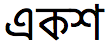} & One hundred & ekSho \\
2 & \includegraphics[width=0.05\textwidth]{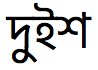} & Two hundred & duiSho \\
3 & \includegraphics[width=0.05\textwidth]{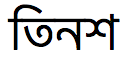} & Three hundred & tinSho \\
4 & \includegraphics[width=0.05\textwidth]{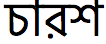} & Four hundred & charSho \\
5 & \includegraphics[width=0.05\textwidth]{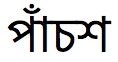} & Five hundred & pachSho \\
6 & \includegraphics[width=0.05\textwidth]{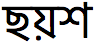} & Six hundred & choySho \\
7 & \includegraphics[width=0.05\textwidth]{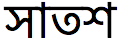} & Seven hundred & satSho \\
8 & \includegraphics[width=0.05\textwidth]{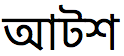} & Eight hundred & aTSho \\
9 & \includegraphics[width=0.05\textwidth]{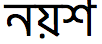} & Nine hundred & noySho \\
10 & \includegraphics[width=0.05\textwidth]{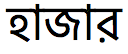} & Thousand & hajar \\
11 & \includegraphics[width=0.05\textwidth]{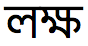} & Lakh & lokSho \\
12 & \includegraphics[width=0.05\textwidth]{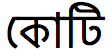} & Crore & koTi \\
13 & \includegraphics[width=0.05\textwidth]{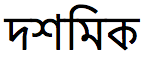} & Decimal & doShomik \\ \hline
\end{tabular}
\end{table}
\paragraph{} The main motivation behind this work was to develop a balanced corpus, meaning that all unique words would occur with  more or less an equal probability. This will be useful for any learning method modeling the speech samples, but this may be a shortcoming in terms of representing the actual prior probabilities of occurring the individual words. The prior probabilities were not considered because we did not have any authentic information about them.

\paragraph{} The rest of the paper is organized as follows. In Section \ref{sec:methodology}, we describe how we have developed the corpus. In the same section we also describe some key features of our corpus. In the next section, Section \ref{sec:analysis}, we evaluate two methods of automatic speech recognition on our corpus. And finally, in Section \ref{sec:conclusion}, we draw the conclusion to our paper.

    \section{Development of Bengali Real Number Speech Dataset}
    \label{sec:methodology}
    
In this section, we give a detailed description of our development process and some key features of it.

\subsection{Data Preparation}
We developed the corpus in two steps. At first, we had to decide what real numbers should be present in the dataset. Therefore, we generated individual random strings containing the Bengali real numbers. We had also generated some strings that did not contain the real numbers, rather, they only contained sequences of numbers (like, ``One Two Three ... Ten''). The algorithm of generating the strings is given in Algorithm \ref{alg:string-generation}.

\begin{algorithm}
\caption{String generation algorithm}
\label{alg:string-generation}
\begin{algorithmic}[1]
\Procedure{StringGeneration}{}
\State $w_1 \gets \text{list of each word for numbers from 0 to 99}$ \Comment{size = 101}
\State $w_2 \gets \text{list of each word for numbers \{100,200,300, ..., 900\}}$ \Comment{size = 9}
\State $w_3 \gets \{$\scalerel*{\includegraphics{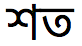}}{B},
\scalerel*{\includegraphics{./img/hajar}}{B}, 
\scalerel*{\includegraphics{./img/lokkho}}{B}, 
\scalerel*{\includegraphics{./img/koti}}{B} \} \Comment{\{hundred, thousand, lakh, crore\}}
\State $w_4 \gets $ \{ \scalerel*{\includegraphics{./img/doshomik}}{B} \} \Comment{\{decimal\}}
\State $N \gets \text{total number of strings to generate}$
\State $counter \gets 0$
\State $S \gets \{\}$ \Comment{final set of generated strings to return}
% \BState \emph{top}:
% \If {$i > \textit{stringlen}$} \Return false
% \EndIf
% \State $j \gets \textit{patlen}$
\State \emph{loop\_outer}:
\State\hspace{\algorithmicindent} $L \gets rand(2,4)$ \Comment{a random integer in range [2,4]} 
\State\hspace{\algorithmicindent} $L \gets L \times 2$ \Comment{the string will have 4 to 8 words}
\State\hspace{\algorithmicindent} $i \gets 0$
\State\hspace{\algorithmicindent} $s \gets ``"$ \Comment{generated string in this iteration}
\State\hspace{\algorithmicindent} \emph{loop\_inner}:
\State\hspace{\algorithmicindent} \hspace{\algorithmicindent} $i \gets i + 2$
\State\hspace{\algorithmicindent} \hspace{\algorithmicindent} $r_1 \gets rand(1,14)$
\State\hspace{\algorithmicindent} \hspace{\algorithmicindent} \textbf{if} {$r_1 > 1$} \textbf{then}
\State\hspace{\algorithmicindent} \hspace{\algorithmicindent} \hspace{\algorithmicindent} $s \gets s + w_1[rand(1,len(w_1))]$
\State\hspace{\algorithmicindent} \hspace{\algorithmicindent}  \textbf{else}
\State\hspace{\algorithmicindent} \hspace{\algorithmicindent} \hspace{\algorithmicindent} $s \gets s + w_2[rand(1,len(w_2))]$
\State\hspace{\algorithmicindent} \hspace{\algorithmicindent} \textbf{end if}
\State\hspace{\algorithmicindent} \hspace{\algorithmicindent} $r_2 \gets rand(1,7)$
\State\hspace{\algorithmicindent} \hspace{\algorithmicindent} \textbf{if} {$r_2 \leq 6$} \textbf{or} $i = L$ \textbf{then} \Comment{the utterance should not end in `decimal'}
\State\hspace{\algorithmicindent} \hspace{\algorithmicindent} \hspace{\algorithmicindent} $s \gets s + w_3[rand(1,len(w_3))]$
\State\hspace{\algorithmicindent} \hspace{\algorithmicindent}  \textbf{else}
\State\hspace{\algorithmicindent} \hspace{\algorithmicindent} \hspace{\algorithmicindent} $s \gets s + w_4[rand(1,len(w_4))]$
\State\hspace{\algorithmicindent} \hspace{\algorithmicindent} \textbf{end if}
\State\hspace{\algorithmicindent} \hspace{\algorithmicindent} $\text{\textbf{if} } i < L$ \textbf{then}
\State\hspace{\algorithmicindent} \hspace{\algorithmicindent}\hspace{\algorithmicindent} \textbf{goto} \emph{loop\_inner}
\State\hspace{\algorithmicindent} \textbf{add} $s$ to $S$
\State\hspace{\algorithmicindent} $counter \gets counter + 1$
\State \hspace{\algorithmicindent} \textbf{if} {$counter < N$} \textbf{then}
\State\hspace{\algorithmicindent} \hspace{\algorithmicindent} \textbf{goto} \emph{loop\_outer}
\State \hspace{\algorithmicindent} \textbf{end if}
\State \textbf{return} $S$
\EndProcedure
\end{algorithmic}
\end{algorithm}

\paragraph{} In the algorithm, we made two clusters (i.e., $\{w_1, w_2\}$ and $\{w_3, w_4\}$) to comprehend all the words in the vocabulary. We constricted the length of the strings to be either four, or six, or eight. A string can be logically split into a number of groups (depending on the length of the string) where each group has exactly two words. The first word comes from either $w_1$ or $w_2$ and the last word comes from either $w_3$ or $w_4$. This reflects the real world scenario. The word `decimal' does not appear as the final word in the sentence, so we have restricted it from appearing as the final word in line number 23 in the algorithm. Thus, we have generated the strings for our corpus. All the generated strings are grammatically valid, but still, there can be a very few strings which are semantically invalid. For example, a semantically invalid string may be like: `\includegraphics[width=0.05\textwidth]{./img/tin-sho} \scalerel*{\includegraphics[]{./img/shoto}}{B} \scalerel*{\includegraphics[]{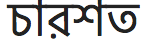}}{B}'. For such strings, we have also incorporated them in our corpus as these strings should not impede the learning process of a recognizer.

\begin{figure}[h]
\centering
\includegraphics[width=1\textwidth]{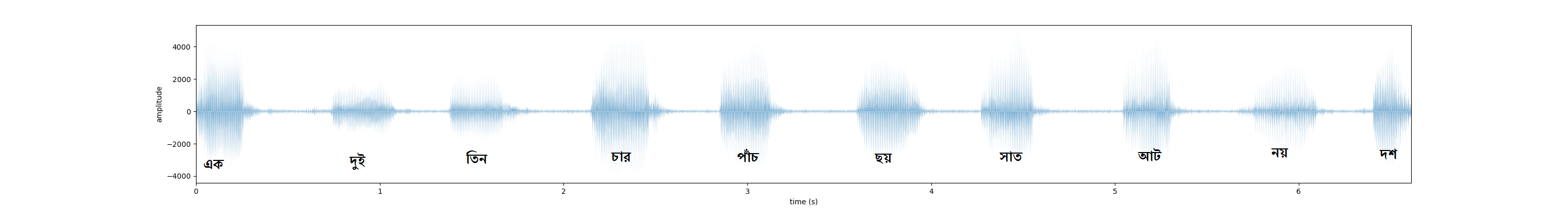}
\caption{Contents of file \texttt{speaker\_1\_1.wav}}
\label{fig:contents-speaker-1-1}
\end{figure}

\paragraph{} After generating the strings, a set of different recording environments was set. This included lab rooms,  class rooms, closed rooms, and etc. The noise in the environment was kept to a minimum level. The volunteers were given the scripts (which contained the randomly generated strings) and were asked to read out in front of a microphone. Their speeches were recorded and filtered using a band-pass filter with the cutoff frequencies 300 Hz and 3 KHz. Finally, we stored the filtered speech samples in ``.wav'' file format with the original bit rate of 256 kbps. A sample speech, in its final representation, is shown in Figure \ref{fig:contents-speaker-1-1}.
%File format: .wav 
%Bit rate: 256kbps
%filtered the speech signal using band pass filter which cutoff frequencies are 300Hz and 3000Hz.

\subsection{Data Organization}
\subsubsection{Transcripts of the Speech Samples}
\label{sec:transcripts-of-speech-samples}
Although we have mentioned before that we had not taken actual prior probabilities into account, but still we made sure that any impossible combination of the words does not occur. For example, the word \scalerel*{\includegraphics{./img/lokkho}}{B} never occurs at the beginning of any speech sample throughout the corpus.
\begin{figure}[h]
\centering
\includegraphics[width=\textwidth]{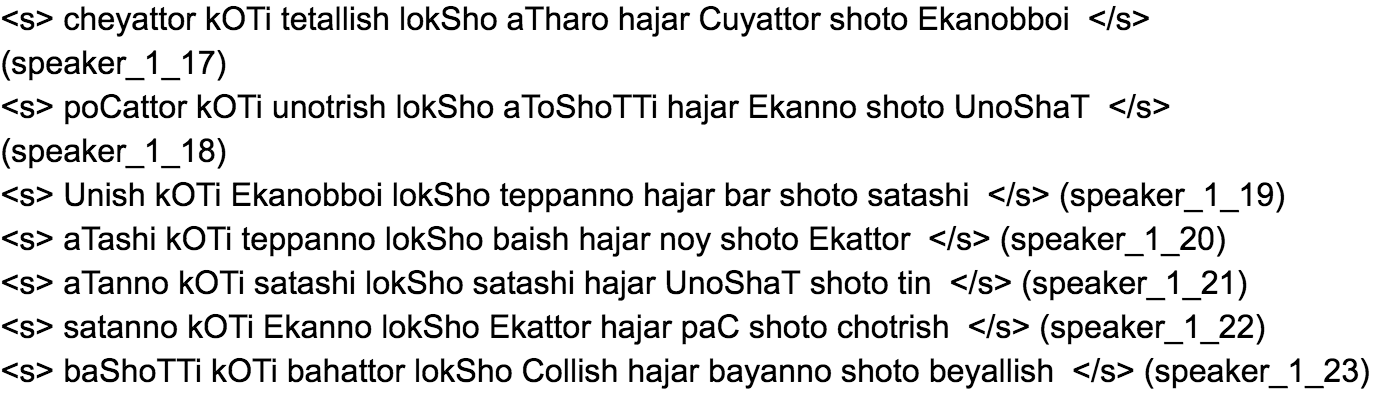}
\caption{Transcript file: \texttt{text-data.txt}}
\label{fig:transcript-file}
\end{figure}
Figure \ref{fig:transcript-file} shows a snapshot of the contents of \texttt{text-data.txt} file of the corpus. This file contains transcripts of all the speech samples. Each line contains the transcript of one speech sample. Each line has mainly two important parts. The actual content contained in the audio file is transcribed within the \texttt{<s></s>} tag as shown in the figure. The second part contains a unique identifier in the format: \texttt{speaker\_\{1-10\}\_\{1-\}}. The first two parts of the identifier in this format describe whose voice it is (i.e., there are ten  speakers in the corpus) whereas the remaining part denotes the recorded file number of that speaker. For example, the second line of Figure \ref{fig:transcript-file} transcribes the speech (of the first speaker) contained in the seventeenth file. This text file may be used as the labels for the speech samples.

\subsubsection{Dictionary of Words to Phonemes}
Individual words (i.e., 115 unique words) were mapped to a set of phonemes according to the rules prescribed in a file ``\texttt{asr\_avro.dic}". These rules follow the popular avro phonetic method. We recognized thirty individual Bengali phonemes which were necessary for constructing individual words in the speech samples. A few examples from the dictionary of matches of the individual words to their corresponding phonemes are given in Table \ref{table:word-to-phone-dictionary}.
\begin{table}
\centering
\caption{Mapping of the words to the phonemes}
\label{table:word-to-phone-dictionary}
\begin{tabular}{|l|l|c|}
\hline
\multicolumn{1}{|c|}{Bengali word} & \multicolumn{1}{|c|}{Corresponding phonemes} & \multicolumn{1}{|c|}{English correspondence} \\ \hline
COUShoTTi & C OU SH O T T I & 64 \\
Ek & E K & 1 \\
Ekotrish & E K O T R I SH & 31 \\
UnoShaT & U N O SH A T & 59 \\
choy & CH O Y & 6 \\
hajar & H A J A R & Thousand \\
poyCollish & P OY C O L L I SH & 45 \\
poytallish & P OY T A L L I SH & 45 \\
teish & T E I SH & 23 \\
teppanno & T E P P A N N O & 53 \\ \hline
\end{tabular}
\end{table}

\paragraph{} This phoneme list is only a supplementary to our data corpus; one can have their own list of phonemes to interpret the sentences corresponding to the speech samples. In our interpretation of the phonemes, the data corpus contains a total of 126,776 phonemes. The frequency of each of the phonemes is given in Table \ref{table:phoneme-frequency}.

\begin{table}
\centering
\caption{The phoneme list and their frequency in the corpus}
\label{table:phoneme-frequency}
\begin{tabular}{l|l|c}
\hline
\multicolumn{1}{c|}{Phonemes} & Occurrences & Normalized frequency (\%) \\ 
\hline \hline
A & 13,943 & 10.80 \\
AI & 193 & 0.15 \\
B & 4,452 & 3.51 \\
C & 3,228 & 2.55 \\
CH & 950 & 0.75 \\
D & 491 & 0.39 \\
E & 7,238 & 5.71 \\
G & 419 & 0.33 \\
H & 12,212 & 9.62 \\
I & 8,603 & 6.76 \\
J & 1,716 & 1.35 \\
K & 4,931 & 3.89 \\
KH & 1,644 & 1.30 \\
L & 3,715 & 2.93 \\
N & 4,477 & 3.53 \\
NG & 307 & 0.24 \\
O & 11,546 & 9.11 \\
OI & 1,357 & 1.07 \\
OU & 344 & 0.27 \\
OY  & 623 & 0.49 \\
P & 4,078 & 3.22 \\
R & 7,740 & 6.11 \\
S & 8,235 & 6.50 \\
SH & 9,152 & 7.22 \\
T & 11,646 & 9.19 \\
TH & 205 & 0.16 \\
U & 2,031 & 1.60 \\
Y & 623 & 0.49 \\
YA & 870 & 0.69 \\ \hline
\end{tabular}
\end{table}

\subsubsection{Organization of the Audio Samples}
The speech samples are organized into an appropriate structure. Each speech sample is under a directory whose name is the unique identifier for that speaker. In other words, each speaker has a dedicated directory for all the speech samples attributed to him. The audio files have an ``.wav'' extension. For example, the twenty first speech sample of the fifth speaker has the name ``\texttt{speaker\_5\_21.wav}". The information about the speech contained in the sample is in ``\texttt{text-data.txt}'' file as described in Section \ref{sec:transcripts-of-speech-samples}.

\subsection{Characteristics of the Data Corpus}
\label{sec:characteristics-of-the-data-corpus}
In this section, we discuss some of the important characteristics of the corpus which provide a more important insight to the dataset. The dataset has 2302 examples, and 17,582 words in total. The transcripts to all the speech samples span 155,908 characters (without the spaces). Our data corpus has more longer words (i.e., the words that have seven or more characters comprise 36\% of the whole dataset) as compared to the shorter words (i.e., the words that have three or less characters comprise 24\% of the whole dataset).

\paragraph{} Interestingly, no real number in Bengali language utters phonemes `M', `BH', `GH', `PH' (or `F'). This is the reason why these phonemes have not appeared in Table \ref{table:phoneme-frequency}. Another observation is that there is a significant amount of disproportions in the frequencies of phonemes (e.g., the phoneme `A' has a normalized frequency 10.8\% while the phoneme `AI' has 0.15\%). This is a limitation of our data corpus, and therefore, any method that would train on this dataset has to pay a significant attention to this discrepancy. As we were not biased to any phoneme while building the dataset, yet  such disproportions arose, it can be inferred that the disproportions of the phonemes exist in the language itself.

    \section{Two Speech Recognizers}
    \label{sec:analysis}

\begin{figure}[h]
\centering
\includegraphics[width=0.75\textwidth]{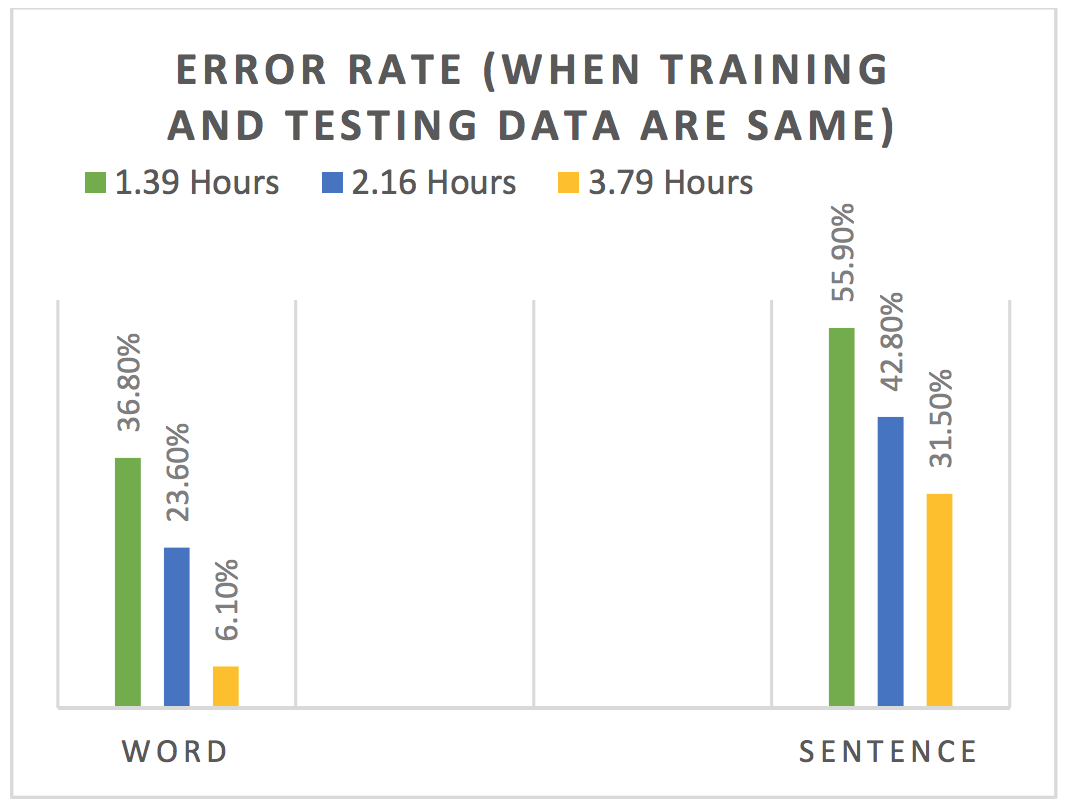}
\caption{Error rates incurred by CMU-Sphinx4 model with the respect to different durations of the corpus. The word error rates are smaller as compared to the sentence error rate because of shorter length dependency modeling. }
\label{fig:different-durations-CMU-Sphinx4}
\end{figure}

In this section, we describe the findings of the experiments undergone with two Bengali speech recognizers \cite{nahid2016noble, nahid2017lstm}. These two recognizers had been developed for recognizing Bengali real numbers. The ideas and methods of these two recognizers are very different.

\paragraph{} To analyze and evaluate these two recognizers we need two types of error metrics; the first one is the word detection error rate ($E_w$) and the second one is the phoneme detection error rate ($E_p$). In our experiment we define them as follows:

\begin{align}
E_p &= \frac{\# \text{mismatches in phonemes}}{\# \text{total phonemes}} \\
E_w &= \frac{\# \text{mismatches in words}}{\# \text{total words}}
\end{align}

\paragraph{} The first recognizer uses a popular recognizer CMU Sphinx 4 \cite{walker2004sphinx} which employs a hidden markov model (HMM). This model has been trained on our corpus with different durations of the dataset and the error rates are shown in Figure \ref{fig:different-durations-CMU-Sphinx4}. These error rates are incurred when the training and the test data are the same. Finally, with different training and test data (with a 80:20 ratio for training and test purposes respectively) this model incurs a 15\% word error rate.

\begin{figure}[h]
\centering
\includegraphics[width=0.75\textwidth]{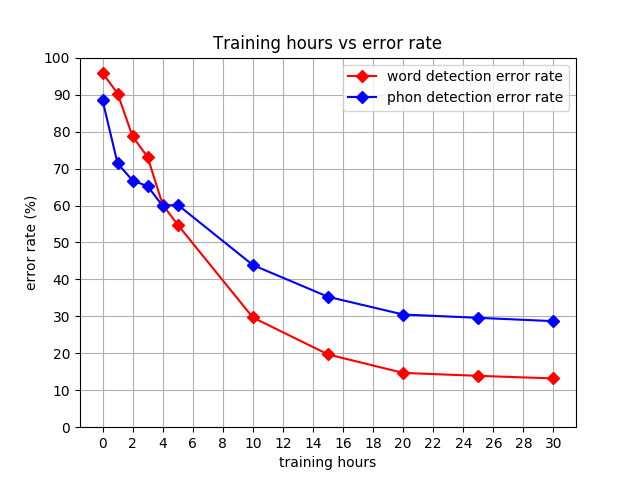}
\caption{The LSTM model's learning process with respect to the length of training period.}
\label{fig:LSTM-hourly-training}
\end{figure}

\paragraph{} The second recognizer tries to recognize the speech samples by dividing them into a number of frames. It is actually a bidirectional long short term memory (LSTM) (a recurrent neural network) that does sequence learning, and has a post-processing algorithm to produce a final output. However, it can only recognize the individual words, not the entire sentence. The hourly training progression is given in Figure \ref{fig:LSTM-hourly-training}. In this figure, the error rates in individual phonemes are also given. The word detection error rate improves after the initial couple of hours because of the post-processing algorithm. Finally, the recognizer incurred a 13.2\% word detection error rate and a 29\% phoneme detection error rate.

    \section{Conclusion}
    \label{sec:conclusion}
    
In this paper, we described the development of Bengali real number audio corpus that comprises of all possible words involving any Bengali real number. The major contribution of our corpus is that it comprehends a great combination of Bengali real numbers. We also provided a brief analysis of our corpus and highlighted some of the key features of it. We have also performed experiments with two methods on our corpus and evaluated their performances. On a future note, we would like to extend our corpus with more combinations of words, and with an increased and balanced number of male and female speakers.

\end{document}